\documentclass[a4paper,11pt,aastex]{article}
\pdfoutput=1 

\usepackage{jcappub} 

\usepackage[T1]{fontenc} 
\usepackage{hhline}
\usepackage{graphicx}
\usepackage{subcaption}
\usepackage{mwe}
\usepackage{comment}
\usepackage[utf8]{inputenc}

\usepackage[toc,page]{appendix}

\title{\boldmath Testing Emergent Gravity on Galaxy Cluster Scales}

\author[1,a]{Andrius Tamosiunas,\note{Corresponding author}}
\author[a]{David Bacon,}
\author[a]{Kazuya Koyama,}
\author[a]{Robert C. Nichol}

\affiliation[a]{Institute of Cosmology and Gravitation, University of Portsmouth,\\Dennis Sciama Building, Burnaby Road, Portsmouth, PO1 3FX, United Kingdom}

\emailAdd{andrius.tamosiunas@port.ac.uk}
\emailAdd{david.bacon@port.ac.uk}
\emailAdd{kazuya.koyama@port.ac.uk}
\emailAdd{bob.nichol@port.ac.uk}

\abstract{Verlinde's theory of Emergent Gravity (EG) describes gravity as an emergent phenomenon rather than a fundamental force.  Applying this reasoning in de Sitter space leads to gravity behaving differently on galaxy and galaxy cluster scales; this excess gravity might offer an alternative to dark matter. Here we test these ideas using the data from the Coma cluster and from 58 stacked galaxy clusters. The X-ray surface brightness measurements of the clusters at $0.1 < z < 1.2$ along with the weak lensing data are used to test the theory. We find that the simultaneous EG fits of the X-ray and weak lensing datasets are significantly worse than those provided by General Relativity (with cold dark matter). For the Coma cluster, the predictions from Emergent Gravity and General Relativity agree in the range of 250 - 700 kpc, while at around 1 Mpc scales, EG total mass predictions are larger by a factor of 2. For the cluster stack the predictions are only in good agreement at around the 1 - 2 Mpc scales, while for $r \gtrsim 10$ Mpc EG is in strong tension with the data. According to the Bayesian information criterion analysis, GR is preferred in all tested datasets; however, we also discuss possible modifications of EG that greatly relax the tension with the data.  
}

\begin{document}
\maketitle
\flushbottom

\section{Introduction}
\label{sec:intro}

Studies in the field of black hole thermodynamics indicate an intimate relationship between thermodynamics and gravity. As shown by Bekenstein (1973), black holes are thermodynamical objects with an entropy proportional to the area of the event horizon \cite{bekenstein}. More recently it was shown that this is a consequence of a more general principle that relates the number of fundamental degrees of freedom in a given region with the surface area associated with the boundary of the region. In black hole physics this sort of reasoning leads to a deep link between the information content of objects that have fallen into a black hole and the quantum fluctuations at the event horizon. This idea is known as the holographic principle, and has intimate connections with the AdS/CFT correspondence; it could be a much more general property of gravitational systems and could shine light on some general features of quantum gravity. 

The recent proposal by Verlinde \cite{a,b} combines these ideas in an attempt to describe gravity as an emergent force rather than a fundamental interaction. This reasoning builds on previous work like \cite{jacobson} and \cite{padmanabhan}, where the Einstein field equations are derived from the area law of entropy. In \cite{a} Verlinde demonstrates a similar result, where Newton's laws of gravity as well as Einstein's field equations are derived starting from the holographic principle. One of the main results in \cite{a} shows how changes in the entropy of a gravitational system can be related to the changes in the gravitational potential acting on a test mass near a spherical mass distribution enclosed by a holographic screen (see fig. 3 in \cite{a}):

\begin{equation}
    \frac{\Delta S}{n} = -k_{B}\frac{\Delta \Phi}{2c^{2}},
\end{equation}

\noindent where $\Delta S$ is the change in entropy, $n$ is the number of bits of information stored on the holographic screen bounding the system, $\Delta \Phi$ is the change in the gravitational potential and $k_{B}$ and $c$ are the Boltzmann constant and the speed of light. This result, (also extended for general mass distributions in \cite{a}) illustrates one of the main principles of Emergent Gravity -- the changes in gravitational potentials are equivalent to the changes of the entropy of the system and in turn the dynamics of the microscopic degrees of freedom of the system. 

The ideas above are easiest to derive and understand in anti-de Sitter space. More recently, in \cite{b}, Verlinde extended the Emergent Gravity formalism to de Sitter space in the hope of applying the results in more realistic cosmological situations. This requires taking into account the entropy and temperature associated with the cosmological horizon. This, in turn, leads to a volume contribution to the usual area law of entropy, i.e. now the entropy of a system on certain scales depends not only on the surface area of the boundary, but also on the volume of the system. In other words, in the Emergent Gravity formalism, instead of directly modifying the Einstein field equations (by adding extra fields, additional dimensions etc.), the entropy area law is modified in the de Sitter space, by adding a volume dependent term. The modified entropy law can then be used to derive the field equations, which will, in general, be different from GR (note, however, that in \cite{b} the main predictions are obtained from general arguments without deriving the field equations). In this way, Emergent Gravity offers a unique approach of modifying the laws of gravity through changing the entropy law.

In \cite{b} it is shown that introducing a central baryonic mass distribution in de Sitter space results in the reduction of the total entanglement entropy of the system, which is equivalent to extra gravitational effects (i.e. a force pointing towards the matter distribution (see Fig. 1.)). These extra gravity effects are comparable in size to the effects usually associated with those of cold dark matter in the standard model of cosmology. 

The entropy change in a spherical system caused by introducing a spherical central distribution of baryonic matter $M_{B}$ can be expressed through the displacement field $u(r)$:

\begin{equation}
    S_{M}(r) = \frac{u(r)A(r)}{V_{0}^{*}}
    \quad\mathrm{with}\quad
    V_{0}^{*}= \frac{2G \hbar}{cH_{0}},
\end{equation}

\noindent where $S_{M}$ is the amount of displaced entropy, $A(r) = 4 \pi r^{2}$ is the surface area of the system, $G$ is the gravitational constant, $c$ is the speed of light, $\hbar$ is the reduced Planck constant and $H_{0}$ is the current value of the Hubble parameter.

To fully describe the entropy displacement by baryonic matter, Verlinde compares the effects described above to the effects of inclusions in elastic materials as described by the linear theory of elasticity. In general, introducing inclusions into elastic materials causes strain $\epsilon$, which can be related to the changes in entropy of the system. In \cite{b} Verlinde notices that the effects of inclusions in elastic materials share certain similarities with the effects of baryonic matter distributions on the entanglement entropy in de Sitter space. An elasticity/gravity correspondence\footnote{For a better understanding of this correspondence see table 1 in \cite{b}.} is then established to derive the exact result for the extra gravitational effects due to entropy displacement. In particular, if the spacetime in our system is treated as an incompressible elastic medium, the strain $\epsilon_{D}(r) = u'(r)$ caused by the baryonic matter is then given by: 

\begin{equation}
\int^{r}_{0}\epsilon_{D}^{2}(r')A(r')dr' = V_{M_{B}},
\label{strain1}
\end{equation}

\noindent where $A$ is the area of a sphere we are integrating over and $V_{M_{B}}$ is a quantity related to the amount of entropy displaced by the baryonic matter distribution $M_{B}$\footnote{Note that $V_{M_{B}}$ is equal to the volume that would contain the amount of entropy that is removed by a mass $M_{B}$ inside a sphere of radius $r$, if that volume was filled with the average entropy density of the universe (see \cite{brouwer} for a wider discussion).} and is given by eq. (\ref{volume}). In \cite{b} it is shown that in de Sitter space $S_{M}(r) = (-2 \pi Mr)/\hbar$, which leads to $\epsilon_{D}(r)$ being given by: 

\begin{equation}
\epsilon_{D}(r) = \frac{8 \pi G}{cH_{0}}\frac{M_{D}(r)}{A(r)},
\label{strain}
\end{equation}

\noindent where $M_{D}(r)$ refers to the ``apparent dark matter'' distribution (defined in eq. \ref{EG})\footnote{Here we want to emphasize that in Emergent Gravity, there is only baryonic matter. However, gravity acts differently on large scales, which can be modeled as a consequence of an effective extra mass distribution, here called $M_{D}$. The effects of $M_{D}$ can then be compared against those of dark matter in standard cosmology.}. The $V_{M_{B}}$ term is given by:


\begin{equation}
V_{M_{B}} = \frac{8 \pi G}{3cH_{0}}M_{B}(r)r.
\label{volume}
\end{equation}

Substituting eq. (\ref{volume}) and (\ref{strain}) into (\ref{strain1}) and integrating, leads to the main result, which we test in this work:

\begin{equation}
M_{D}^{2}(r) = \frac{cH_{0}r^{2}}{6G} \frac{d(M_{B}(r)r)}{dr},
\label{EG}
\end{equation}

\noindent where $M_{D}(r)$ is the apparent dark matter mass enclosed in $r$ and $M_{B}(r)$ is the baryonic mass. This can be interpreted as an effective dark matter distribution caused by gravity acting differently on large scales, rather than a new form of matter as in the $\Lambda$CDM framework. Hence Verlinde's Emergent Gravity offers an alternative solution to the problem of dark matter. 

This result has a number of interesting consequences. For instance, computing the total acceleration due to $M_{B}$ and $M_{D}$, assuming that the baryonic mass is concentrated in the centre, leads to the result below that agrees well with the baryonic Tully-Fisher relation:

\begin{equation}
\frac{GM_{D}(r)}{r^{2}} = \sqrt{\frac{a_{0}GM_{B}(r)}{6r^{2}}}, 
\end{equation}

\noindent with $a_{0} = cH_{0}$ (the scale familiar from modified Newtonian dynamics \cite{mond1})\footnote{Note that there is an ongoing debate whether the derivation of this result is self-consistent (see \cite{stojkovic1} for the full discussion).}. Similarly, applying eq. (\ref{EG}), for extended mass distributions in galaxy clusters, highly reduces the missing mass problem, hence possibly offering an alternative to dark matter on galaxy and cluster scales.

It is important to note that the equations outlined above will be valid under a certain set of assumptions that are outlined below:

\begin{itemize}
    \item Emergent Gravity predictions are only applicable for approximately spherically symmetric, sufficiently isolated and non-dynamic mass distributions. This means that, for example, the Bullet cluster would not be a valid test case for Emergent Gravity. 
    \item Since there is no rigid description of cosmology in Verlinde's Emergent Gravity yet, all the equations are only valid for the current value of the Hubble parameter, $H_{0}$ (i.e. $H(z)$ will be approximated as $H_{0}$ and only small redshift clusters will be considered). This also implies that Verlinde's theory is not capable of addressing such phenomena as the CMB power spectrum and structure formation. However, note that more recent Emergent Gravity approaches, such as Hossenfelder's covariant approach (see \cite{hossenfelder}) could in principle address the CMB.  
    \item There is also no geodesic equation in Emergent Gravity so far, so we will make a crucial assumption that lensing works in EG the same way as in GR. In particular, following the work in \cite{brouwer}, we will assume that the extra gravity effects predicted by Emergent Gravity, affect the paths of photons in the same way as dark matter does in GR. Future theoretical and observational work will be required to test the validity of this assumption; however, in this work we assume that it is valid and test the outcomes of such an assumption. 
    \item As discussed by Brouwer et al. (2016), the effects of EG are only expected to become important in the regime where the volume law contribution of entropy ($S \propto V$) is significantly larger than the entropy displaced by baryonic matter, $M_{B}$. This, following eq. 18 in \cite{brouwer}, is expressed by introducing a minimal radius, $r_{min}$, above which we expect the EG effects to become noticeable, as described by the following inequality:

\begin{equation}
r > \sqrt{\frac{2M_{B}(r)G}{cH_{0}}}.
\label{rmin}
\end{equation}
\end{itemize}

\begin{figure}
  \centering
    \includegraphics[width=0.35\textwidth]{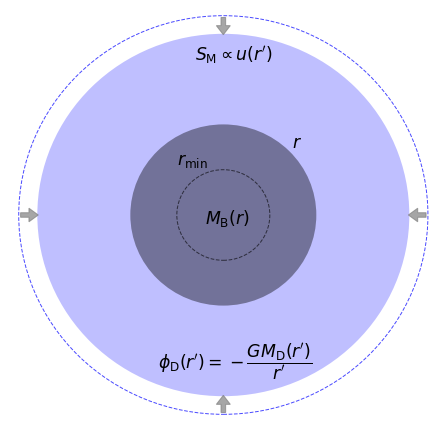}
    \caption{Diagram illustrating the physical system that was tested. Introducing a central baryonic distribution $M_{B}$ (for instance a galaxy cluster here shown in dark purple) causes a reduction of the entanglement entropy $S_{M}$, which is quantified by the displacement field $u(r)$. This results in a central force and a potential $\phi_{D}(r')$, which can be calculated using the scaling relation \ref{EG}. These effects are expected to become significant at radii larger than $r_{min}$.  }
\end{figure}

The theory has already been tested with several methods and on a range of scales. In \cite{brouwer} the average surface mass density profiles of isolated central galaxies were used to test Emergent Gravity, finding that the predictions of the theory are in good agreement with measured galaxy-galaxy lensing profiles. More recently the predictions of Emergent Gravity were compared with the predictions from selected modified gravity theories for a single galaxy cluster, showing that Emergent Gravity only approaches the measured acceleration profile data at the outskirt of the cluster \cite{hodson}. Similarly, the X-ray and weak lensing data from the A1689 cluster 
was used in \cite{zwicky_EG} to compare the predictions of Emergent Gravity with some selected modified gravity models, finding that EG fails to account for the missing mass in the mentioned cluster. Most recently in \cite{halenka_miller} the authors have used mass densities of a sample of 23 galaxy clusters to test the predictions of EG on galaxy cluster scales. They found that EG could only correctly predict the baryon and dark matter mass profiles at around the virial radius, while being ruled out at $>5\sigma$ level in the other parts of the clusters. However, as the authors pointed out, fully accounting for the systematic uncertainties and modifying certain assumptions in the model leads to a good agreement between GR and Emergent Gravity (see the appendix \ref{appendix:c}). 

In this work we will test the scaling relation (\ref{EG}) on galaxy cluster scales using a combination of observational probes. We will start by testing the EG predictions using the temperature and weak lensing profile data for the Coma cluster. Then we will apply a similar method to test EG using stacked galaxy clusters. The following cosmology was used to calculate the GR results: $\Omega_{m} = 0.27$, $\Omega_{\lambda} = 0.73$ and $H_{0} = 70$km s$^{-1}$ Mpc$^{-1}$.

\section{Testing Emergent Gravity with the Coma Cluster}

Galaxy clusters, being the largest gravitationally bound systems, offer a natural setting for testing models of gravity. Having regions of high and low density as well as a mass distribution dominated by dark matter, clusters have been used extensively for testing models with screening mechanisms and comparing the predictions with General Relativity. In this work we use an approach similar to the one developed by Terukina et al. (2014) and Wilcox et al. (2015), who tested $f(R)$ gravity with chameleon screening in the Coma cluster as well as a 58 cluster stack \cite{Terukina, Wilcox}. In these works, multiple probes are used to constrain the modified effects of gravity on the outskirts of galaxy clusters under the assumption of hydrostatic equilibrium. Here we use the intracluster gas temperature profiles along with the X-ray surface brightness data to determine the baryonic mass distribution in a given cluster and to calculate the predicted weak lensing signal, which is then compared with the actual weak lensing data. The same procedure is done for the model of Emergent Gravity and the standard model (GR + cold dark matter described by an Navarro-Frenk-White profile) and the results are then compared. The data used includes the profiles from the Coma cluster as well as the stacked profiles for 58 galaxy clusters as described in \cite{Wilcox}.

The Coma cluster (Abel 1656) is a large well-studied nearby ($z =0.0231$) galaxy cluster with over 1,000 identified galaxies. The cluster has an extensively-studied mass distribution and has been the subject of numerous weak lensing and X-ray studies. In the equations below we illustrate how the temperature profile of the Coma cluster can be used to determine the total mass distribution and, in turn, to calculate the predicted weak lensing signal.

Assuming hydrostatic equilibrium we can relate pressure to the mass distribution through:

\begin{equation}
\frac{1}{\rho_{gas}(R)} \frac{dP_{gas}}{dr}  = -\frac{GM(<r)}{r^{2}},
\end{equation}

\noindent where $\rho_{gas}$ is the gas density, $P_{gas}$ is the total pressure and $M(<r)$ is the mass enclosed in radius $r$. This allows us to calculate the gas temperature by noting that $P_{thermal} = n_{gas}kT_{gas}$. Integrating eq. (2.1) gives: 

\begin{equation}
T_{gas}(r) =    -\frac{m_{p}\mu}{n_{e}(r)k}\Big( \int^{r}_{0} n_{e}(r') \frac{GM(<r')}{r'^{2}}  + P_{gas,0} \Big) dr',
\label{t_gas}
\end{equation}

\noindent where $\mu$ is the mean molecular weight, $m_{p}$ is the proton mass, $n_{e}(r)$ is the electron number density and the last term, the central pressure, is an integration constant. For a fully ionised gas, the mean molecular weight is defined by $\mu(n_{e} + n_{H} + n_{He})m_{p} = m_{p}n_{H} + 4m_{p}n_{He}$ with $n_{e} = n_{H} + 2n_{He}$, where $n_{e}$, $n_{H}$ and $n_{He}$ are the electron, hydrogen and helium number densities respectively. Assuming the mass fraction of hydrogen of $n_{H}/(n_{H} + 4n_{He}) = 0.75$ (following previous work as described in \cite{Terukina} and \cite{Wilcox}) leads to $\mu = 0.59$. Here we also expressed the gas density explicitly by noting that $\rho_{gas} = \mu m_{p}n_{gas}$ and that $n_{e} = n_{gas}(2 + \mu)/5$. Eq. (\ref{t_gas}) allows us to calculate the gas temperature in the Coma cluster, given that we have a way to measure the mass distribution $M(<r)$. In this work we adopted the standard beta-model electron density profile (e.g. Cavaliere $\&$ Fusco-Femiano (1976)) \cite{fusco_femiano}. 




The baryonic mass distribution is given by:

\begin{equation}
M_{B}(<r) = M_{gal}(<r) + 4 \pi m_{a} \int^{r}_{0} n_{e}(r') r'^{2}dr',
\label{baryonic_mass}
\end{equation}

\noindent where we summed the total stellar galaxy mass with the intracluster gas mass and $m_{a}$ is the average mass of an atom in the cluster gas, given by $2m_{H}/(1 + X)$ where $m_{H}$ is the Hydrogen mass and $X$ is the mass fraction of the Hydrogen atoms.

In order to estimate the galaxy mass distribution in the Coma cluster, we queried the SDSS data catalogue (Data Release 14) for the median estimate of the total stellar masses of galaxies located within the 180 arcminute diameter around the central point of the cluster for $0.01<z<0.05$ \cite{sdss}. This region was then split into radial bins of 5 arcminutes, and for each cylindrical shell we summed the stellar masses for all the detected galaxies. This results in a galaxy mass distribution in a spherical region of $r \simeq 1.4$ Mpc around the centre of the cluster. Summing the stellar galaxy and the X-ray emitting gas mass distributions gives us a good measure of the total baryonic mass distribution, which can then be used to calculate the total mass distribution using eq. (\ref{EG}). Finally, having the total mass distribution for the cluster, we have all that is needed to compute the weak lensing predictions.

In order to compare the predictions from Emergent Gravity with those from standard cosmology (GR + dark matter), we chose to describe the dark matter distribution in the cluster by the NFW profile:

\begin{equation}
M_{NFW}(<r) = 4 \pi \rho_{s} r_{s}^{3} \Big( \ln(1 + r/r_{s}) - \frac{r/r_{s}}{1+r/r_{s}} \Big),
\label{nfw1}
\end{equation}

\noindent where $\rho_{s}$ is the characteristic density and $r_{s}$ is the characteristic scale \cite{nfw1}. This was then used to calculate the total mass in the cluster and, in turn, to predict the weak lensing profile. 

Following the approach taken by Brouwer et al. (2016) and using the equations described by Wright et al. (2000), we calculated the weak lensing profiles as follows: \cite{brouwer, nfw1}:

\begin{equation}
\gamma_{t}(r) = \frac{\bar{\Sigma}(r) - \Sigma(r)}{\Sigma_{c}},
\end{equation}

\noindent where $\gamma_{t} (r)$ is the tangential shear, while $\Sigma (r)$ and $\Sigma_{c}(r)$ are correspondingly the surface density and critical surface density (see appendix \ref{appendix:b} for the full expressions). The surface density of a given radial density distribution is given by:

\begin{equation}
\Sigma(r) = \int^{\infty}_{-\infty} \rho(r) dr = \int^{\infty}_{-\infty} \rho(R,z) dz,
\end{equation}

\noindent where we switched to cylindrical coordinates ($R$, $\phi$, $z$) centered on the central point of our cluster. $\Delta \Sigma(R)$ for both baryonic and apparent dark matter can be calculated using the general expression:

\begin{equation}
\Delta \Sigma(R) = \bar{\Sigma}(<R) - \Sigma(R) = \frac{2\pi \int^{R}_{0} R' \Sigma(R') dR'}{\pi R^{2}} - \Sigma(R).
\end{equation}

\noindent In case of Emergent Gravity, the shear equations are then given by:

\begin{equation}
\gamma_{t}(r) = \frac{\Delta \Sigma_{EG}(R)}{\Sigma_{c}} = \frac{\Delta \Sigma_{B}(R) + \Delta \Sigma_{D}(R)}{\Sigma_{c}},
\label{shear}
\end{equation}

\noindent where we have split $\Delta \Sigma$ into contributions from baryonic and apparent dark matter for the surface density.

Having laid out the main equations at this point it is worth noticing that to derive the total mass distribution of the cluster, we need to choose a way of parametrizing the electron number density $n_{e}(r)$. This is done by using the simple isothermal beta profile of the following form: $n_{e} = n_{0}(1 + (r/r_{1})^{2})^{b_{1}}$. The only free parameters for the Emergent Gravity model appearing in the equations above are then $n_{0}$, $r_{1}$, $b_{1}$ and $T_{0}$ (central temperature). On the other hand (given our assumption that dark matter is distributed according to the NFW profile), for the GR model we have the following free parameters: $n_{0}$, $r_{1}$, $b_{1}$ and $T_{0}$, $c_{v}$ and $M_{v}$ (where the last two are the concentration and virial mass parameters). The values for the free parameters were then obtained by looking for solutions that fit the temperature profile data and, at the same time, produce weak lensing predictions which agree well with the observational data (in other words, both datasets were fit simultaneously by minimizing the combined value of $\chi^{2}_{T_{gas}} + \chi^{2}_{\gamma_{t}}$). The data used included the X-ray temperature profile (combined from Snowden et al. (2008) and Wik et al. (2009)) and the weak-lensing profile (Gavazzi et al. (2009), Okabe et al. (2010)) of the Coma cluster \cite{Terukina, snowden, wik, gavazzi, okabe}.

The data fitting was performed by minimizing the combined residuals using the limited memory Broyden–Fletcher–Goldfarb–Shanno (L-BFGS) algorithm available from the SciPy python library \cite{scipy}. The 1-$\sigma$ confidence intervals were determined using the in-built features of the SciPy.optimize library, which use the estimated inverse Hessian matrix to calculate the standard deviation of each best-fit parameter. The $\chi^{2}$ values were calculated using the standard formula: $\chi^{2} = \sum_{i} (C_{i} - O_{i})^{2}/\sigma_{i}^{2}$, where $C_{i}$ refers to the calculated values, $O_{i}$ to the observed values, $\sigma_{i}$ to the variance at a given data point. The covariance matrix here was assumed to be diagonal, however, in the case of the cluster stack data, we used the full covariance matrix (see section \ref{section3} for the full discussion). The best-fit results for the standard model (GR + dark matter) and Emergent Gravity results are summarized in table \ref{table:coma} and figure \ref{fig:main_results}. The goodness of fit statistics are given in table \ref{coma_goodness_of_fit}.

\newcommand\T{\rule{0pt}{3.0ex}}       
\newcommand\B{\rule[-2.0ex]{0pt}{0pt}} 

\begin{table}[]
\noindent\makebox[\textwidth]{%
\begin{tabular}{ccccccc}
\centering
& $\boldsymbol{n_{0}}$ \textbf{($\mathrm{\mathbf{cm}^{-3}}$)} & $\boldsymbol{r_{1}}$ \textbf{(Mpc)} & $\boldsymbol{b_{1}}$  & $\boldsymbol{T_{0}}$ \textbf{(keV)} & $\boldsymbol{M_{v}}$ ($\boldsymbol{\mathrm{M\textsubscript{\(\odot\)}}}$)  & $\boldsymbol{c_{v}}$  \\ \hline \hline
\textbf{GR:}                   &  $4.2^{+0.21}_{-0.17} \times 10^{-3}$ \T \B  &$0.07^{+0.04}_{-0.04}$    & $-0.201^{+0.512}_{-0.512}$    &  $8.77^{+0.60}_{-0.61}$   & $2.39^{+1.18}_{-1.16} \times 10^{14}$                        &     $ 4.68^{+1.37}_{-1.36}$                   \\
\textbf{EG:}        &  $3.2^{+0.21}_{-0.20} \times 10^{-3}$ \T \B   &  $0.26^{+0.026}_{-0.025}$   &  $-0.615^{+0.056}_{-0.062}$   &  $9.18^{+0.13}_{-0.14}$  &            n/a            &         n/a               \\ \hline
\end{tabular}}
\caption{The best-fit parameters for the standard model (GR + dark matter) and the model with gravity behaving according to EG fitted to the Coma cluster data (c.f. figure \ref{fig:main_results}). }
\label{table:coma}
\end{table}


\begin{figure}
\begin{subfigure}{.5\linewidth}
\centering
\includegraphics[width=.9\textwidth]{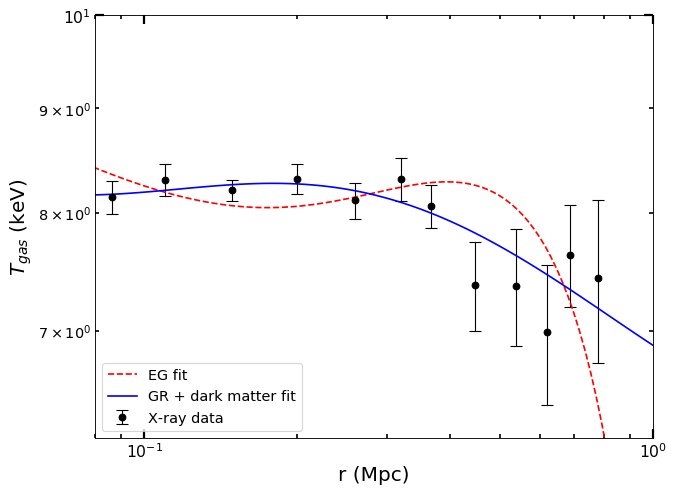}
\caption{\small Gas temperature fit in EG and GR}
\label{fig:mean and std of net14}
\end{subfigure}%
\begin{subfigure}{.5\linewidth}
\centering
\includegraphics[width=.9\textwidth]{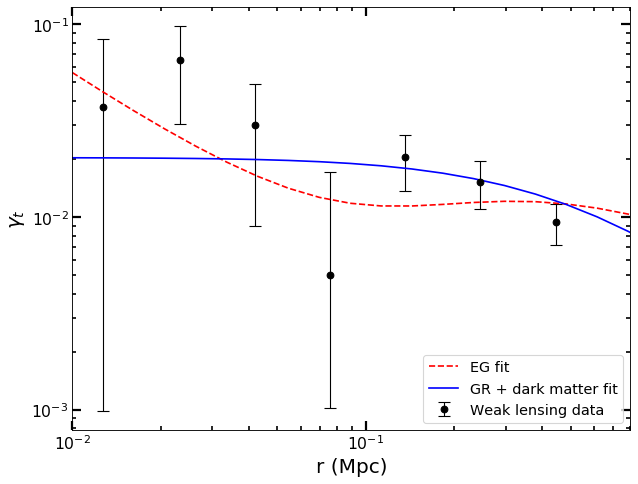}
\caption{\small Weak-lensing fit in EG and GR}
\label{fig:mean and std of net24}
\end{subfigure}\\[1ex]
\begin{subfigure}{0.5\linewidth}
\centering
\includegraphics[width=.9\textwidth]{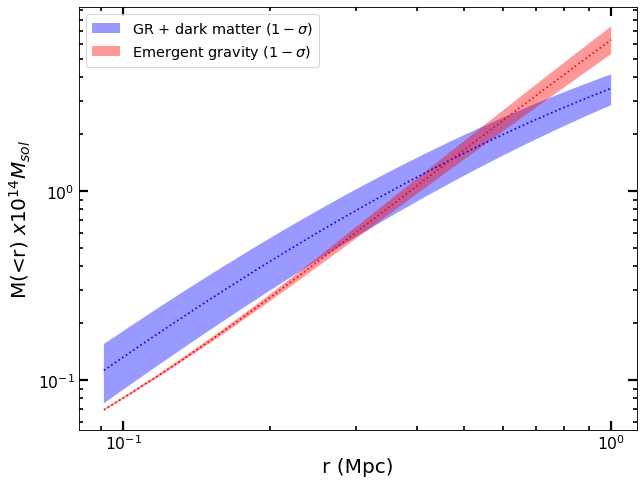}
\caption{\small Total mass profiles in the two models}
\label{fig:fig:mean and std of net44}
\end{subfigure}
\begin{subfigure}{.5\linewidth}
\centering
\includegraphics[width=.9\textwidth]{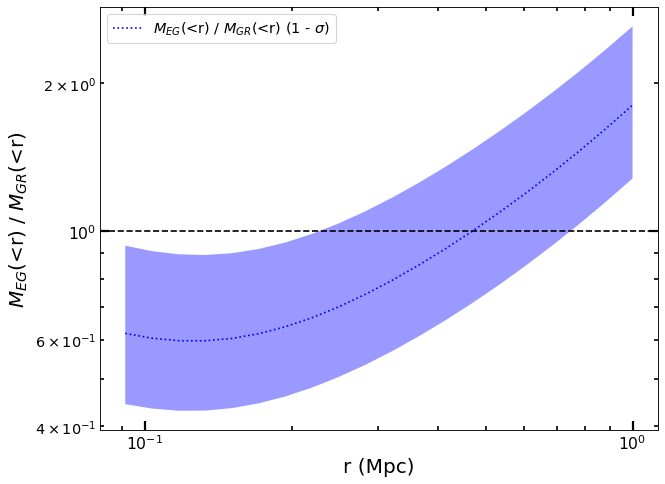}
\caption{\small Ratio of the mass profiles}
\label{fig:mean and std of nets}
\end{subfigure}

\caption{A comparison of the Emergent Gravity and the standard model (GR + dark matter described by the NFW profile) results. Figure $a$ shows the gas temperature fit for both models. Figure $b$ shows the weak lensing fits. Figure $c$ shows the mass distributions calculated using the best-fit parameters for both models (with the contours corresponding to the 1-$\sigma$ confidence intervals). Figure $d$ shows the ratio of the two mass distributions. The goodness of fit statistics are summarized in table \ref{coma_goodness_of_fit}. }
\label{fig:main_results}
\end{figure}

Once the baryonic mass distribution was determined, we investigated on what radial scale we expect the EG effects to become noticeable by calculating the minimal radius $r_{min}$ using eq. (\ref{rmin}). This was done in two different ways -- by assuming that the total mass is situated at the centre of the cluster (point mass approximation) and by using the full radial mass distribution, taking into account the galaxy mass and the intracluster gas components. Taking into account the total mass distribution gives a more accurate estimation for $r_{min}$ to be $8.5$ kpc, which is significantly less than $94$ kpc found using the point mass approximation. We expect the Emergent Gravity effects to become significant at scales larger than $r_{min}$, but the shift to the EG regime will not be instantaneous; however these values show that the scales in our dataset are well within the region where we expect deviations from GR to be detectable.  

\begin{table}[]
\centering
\begin{tabular}{lcccl}
                    & \textbf{$\chi^{2}$} & \textbf{Data points} & \textbf{D.o.f} & \textbf{BIC} \\ \hline \hline
\textbf{GR:\,\,\,\,\,\,\,\,\,\,\,\,} &   21.8                                                  &    19                    &  6               &   20.3            \\
\textbf{EG:\,\,\,\,\,\,\,\,\,\,\,\,} &      29.9                                                &   19                     &    4            & 21.8              \\ \hline
\end{tabular}
\caption{Goodness of fit statistics. $BIC$ is the Bayesian information criterion statistic. The statistics are combined for both the X-ray temperature and the weak lensing fits.}
\label{coma_goodness_of_fit}
\end{table}

We have calculated the typical values for $r_{min}$ for different mass scales and compared this to the typical sizes of different systems. The results are summarized in table \ref{r_min_table}. For instance, for the Solar System, $r_{min}$ is roughly 100 times larger than the size of the system, hence no effects should be noticeable in the local tests of gravity. This agrees well with the conclusions of \cite{hossenfelder}. On the other hand, we see that EG effects are expected be observable on galaxy and cluster scales.

\begin{table}[h!]
\centering
\begin{tabular}{lllll}
 \textbf{Scale}& \textbf{Typical mass ($\boldsymbol{\mathrm{M\textsubscript{\(\odot\)}}}$)} & \textbf{Typical size (Mpc)}  & \textbf{$\boldsymbol{r_{min}}$ (Mpc)} &  \\ \hline \hline
 Solar system& 1.0014   & $5.8 \times 10^{-10}$ & $2 \times 10^{-8}$  &  \\
 Galaxy& $10^{10} - 10^{11}$ & $3-6 \times 10^{-2}$  & $2-6 \times 10^{-3}$ &  \\
 Galaxy cluster&$10^{14} - 10^{15}$& $2-10$  & 0.2 - 0.64&  \\ 
 Coma cluster & $2.2 \times 10^{13}$ & $6$ & 0.094 &  \\ 
 Cluster stack& $1.3 \times 10^{13}$  &  4  &  0.073  &  \\ \hline
\end{tabular}
\caption{\label{tab:i} Typical sizes and average masses of different objects along with the values of $r_{min}$ assuming the point mass approximation. For the Coma cluster and the cluster stack, we have chosen the mean mass in the region of interest covered by our data rather than the full mass. The typical sizes were chosen in the same manner.}
\label{r_min_table}
\end{table}

\section{Testing Emergent Gravity with Stacked Galaxy Clusters}
\label{section3}

As discussed in \cite{Terukina} there are a number of potential issues with using the Coma cluster for testing gravity, which we need to take into account in order to accurately interpret the results given in the previous section. In particular, the cluster is known to deviate from spherical symmetry and the available data, especially for weak lensing, is fairly limited. To mitigate some of these issues and to test the effects of Emergent Gravity in a larger sample of galaxy clusters we followed a similar approach to that taken in \cite{Wilcox}, where 58 clusters with redshifts ranging between $0.1 < z < 1.2$ were stacked using X-ray (from the XMM Cluster Survey) and weak lensing data (from the Canada France Hawaii Telescope Lensing Survey) \cite{XMM, cfhtlens}. Stacking clusters in such a way averages away most irregularities in shape and density and provides an approximation to an average galaxy cluster. In addition, the signal to noise ratio is improved.  

The cluster stack was produced by Wilcox et al. (2015) by first re-scaling the 58 combined images of individual clusters to a standard projected size. This was done by estimating $M_{200}$, the mass enclosed within a sphere at which the average density is 200 times the critical density, and then calculating the corresponding radius, using the prescription described in Sahlen et al (2009) \cite{sahlen}. The 58 cluster images were then rescaled using linear interpolation to a common 500 by 500 pixel format, where each had an $r_{200}$ radius of 125 pixels. Finally, the cluster images were stacked by taking the mean values across all images of the surface brightness and weak lensing at each pixel. 

In order to determine the galaxy mass distribution for our cluster stack, we queried the CFHTLenS survey catalog \cite{cfhtlens} for each individual cluster following a similar procedure as before for the Coma cluster. In particular, for each cluster the galaxy stellar masses were summed in concentric cylindrical shells. The results were then averaged to determine the mean galaxy mass distribution and the corresponding uncertainty for the cluster stack. 

The cluster stack has a number of important properties in the context of the assumptions under which EG predictions are significant. In particular, most galaxy clusters in the dataset are isolated from other nearby mass distributions (see fig. 4 in \cite{Wilcox}). In addition, our dataset consists of clusters with a mean redshift of $z \approx 0.33$, justifying the assumption that we can neglect the effects of varying Hubble parameter $H(z)$ in our test. Finally, the cluster stack has been binned in terms of temperature, to approximately separate it into galaxy groups and galaxy clusters. This allows us to investigate how well the theory in question works for objects of significantly different masses.  


As before, we have used eq. (\ref{shear}) to calculate the tangential shear profiles. For the X-ray dataset, however, we fit the the projected X-ray surface brightness $S(r_{\perp})$, given by:

\begin{equation}
S(r_{\perp}) = \frac{1}{4\pi (1 + z_{cl})^{4}} \int n_{e}^{2}\Big(\sqrt{r_{\perp}^{2} + z^{2}}\Big) \lambda(T_{gas}) dz, 
\end{equation}

\noindent where $z_{cl}$ is the cluster redshift, $n_{e}(r)$ is the electron number density, $\lambda$ is the temperature-dependent cooling function\footnote{The cooling function was calculated using XSPEC software (Arnaud 1996) and utilised the APEC model (Smith et al. 2001) over a range of 0.5keV to 2keV \cite{arnaud, smith}.} and $r_{\perp}$ and $z$ are the projected radius and redshift from the cluster center correspondingly.


As before, we have the same free parameters $n_{0}$, $r_{1}$, $b_{1}$ and $T_{0}$, which were determined by simultaneously fitting the surface brightness and the weak lensing datasets. The total mass profiles were then calculated using the obtained best-fit parameters. The results were compared with the analogous results calculated in GR (with two extra free parameters $M_{v}$ and $c_{v}$). The best-fit was performed using the non-linear least-squares minimization using the python LmFit library \cite{lmfit}. In particular, the Levenberg–Marquardt algorithm was used to determine the best-fit parameter values and the corresponding confidence intervals. 

To characterize the goodness of fit we followed the approach taken in appendix A in \cite{Wilcox}. In particular, for the weak lensing data we approximated the covariance matrix as diagonal. For the surface brightness data the covariance matrix was included in the $\chi^{2}$ calculations to account for the correlations between the surface brightness radial bins\footnote{To put it simply, the mentioned covariance matrix, $C_{i,j}$ quantifies how the change in an ith data point affects the jth data point.}. The results (split into two temperature bins) are summarized in table \ref{table:stack}. The goodness of fit statistics are summarized in table \ref{cluster_stack_goodness_of_fit}.

\begin{table}[]
\noindent\makebox[\textwidth]{%
\begin{tabular}{ccccccc}
\centering
& $\boldsymbol{n_{0}}$ \textbf{($\mathrm{\mathbf{cm}^{-3}}$)} & $\boldsymbol{r_{1}}$ \textbf{(Mpc)} & $\boldsymbol{b_{1}}$  & $\boldsymbol{T_{0}}$ \textbf{(keV)} & $\boldsymbol{M_{v}}$ ($\boldsymbol{\mathrm{M\textsubscript{\(\odot\)}}}$)  & $\boldsymbol{c_{v}}$  \\ \hline \hline
\textbf{GR} (Bin 1):  &$5.5^{+1.8}_{-1.8} \times 10^{-3}$ \T \B  &$0.023^{+0.009}_{-0.009}$    & $-0.59^{+0.04}_{-0.04}$    &  $7.7^{+5.5}_{-5.3}$   &$4.0^{+2.2}_{-2.0} \times 10^{14}$                        &     $7.00^{+1.37}_{-1.49}$                  \\
\textbf{GR} (Bin 2):        &  $8.9^{+1.8}_{-1.8} \times 10^{-3}$ \T \B   &  $0.021^{+0.008}_{-0.009}$   &  $-0.57^{+0.04}_{-0.04}$   &  $6.6^{+5.5}_{-5.3}$  &           $9.61^{+2.52}_{-2.47} \times 10^{14}$            &         $ 4.95^{+1.42}_{-1.51}$             \\
\textbf{EG} (Bin 1):        &  $5.5^{+0.4}_{-0.3} \times 10^{-3}$ \T \B   &  $0.096^{+0.03}_{-0.02}$    &  $-1.00^{+0.04}_{-0.03}$   &  $8.3^{+0.95}_{-0.92}$  &            n/a            &         n/a               \\ 
\textbf{EG} (Bin 2):        &  $3.2^{+0.4}_{-0.4} \times 10^{-3}$ \T \B   &  $0.062^{+0.03}_{-0.02}$   &  $-0.70^{+0.03}_{-0.03}$   &  $7.0^{+0.93}_{-1.01}$  &            n/a            &         n/a               \\ \hline

\end{tabular}}
\caption{The best-fit parameters for the standard model (GR + dark matter) and the EG model for the 58 cluster stack data. Bins 1 and 2 refer to the $T > 2.5$ keV and $T < 2.5$ keV temperature bins correspondingly. }
\label{table:stack}
\end{table}

\begin{figure}
\begin{subfigure}{.5\linewidth}
\centering
\includegraphics[width=.85\textwidth]{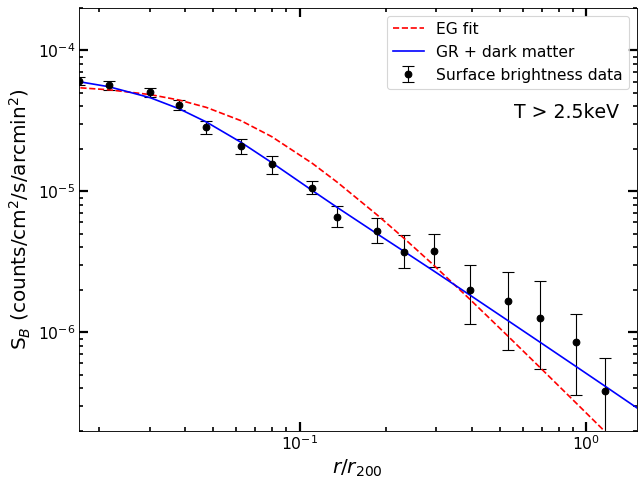}
\caption{X-ray surface brightness fits}
\label{fig:mainA}
\end{subfigure}%
\begin{subfigure}{.5\linewidth}
\centering
\includegraphics[width=.85\textwidth]{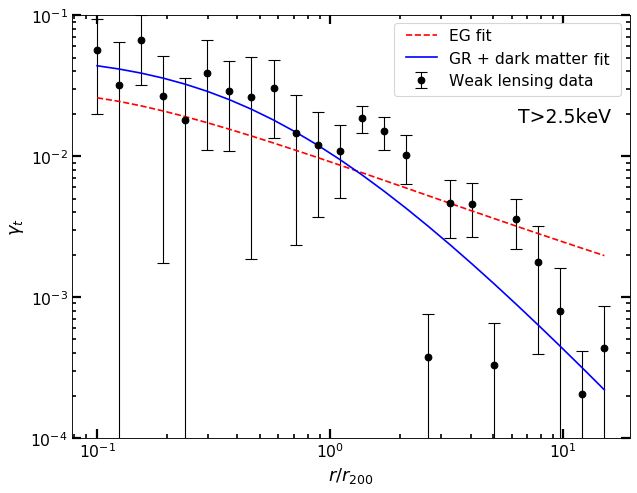}
\caption{Weak lensing fits}
\label{fig:mainB}
\end{subfigure}\\[1ex]
\begin{subfigure}{0.5\linewidth}
\centering
\includegraphics[width=.85\textwidth]{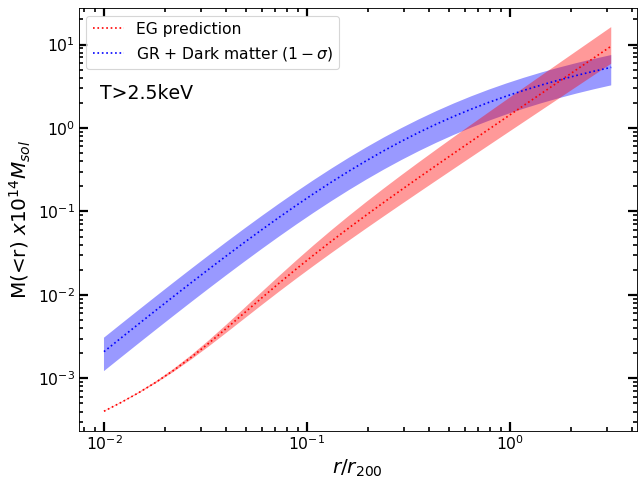}
\caption{Mass distributions in the two models}
\label{fig:mainC}
\end{subfigure}
\begin{subfigure}{.5\linewidth}
\centering
\includegraphics[width=.85\textwidth]{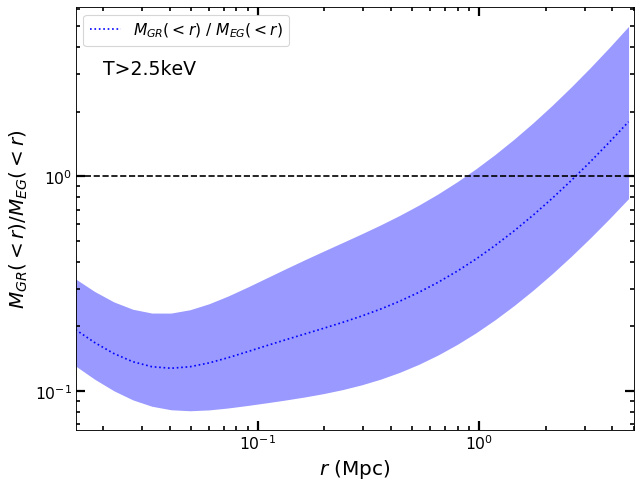}
\caption{Ratio of the mass distributions}
\label{fig:mainD}
\end{subfigure}

\caption{A comparison of the Emergent Gravity and the standard model (GR + dark matter described by the NFW profile) results (for the $T > 2.5$ keV bin roughly corresponding to galaxy clusters). Figure $a$ shows the surface brightness fit for both models. Figure $b$ shows the weak lensing (tangential shear) fit for both models. The mass profiles were calculated using the best-fit parameters, with the blue and red bands corresponding to the 1-$\sigma$ confidence intervals. The goodness of fit statistics are summarized in table \ref{cluster_stack_goodness_of_fit}.}
\label{fig:main_results}
\end{figure}






\begin{figure}
\begin{subfigure}{.5\linewidth}
\centering
\includegraphics[width=.85\textwidth]{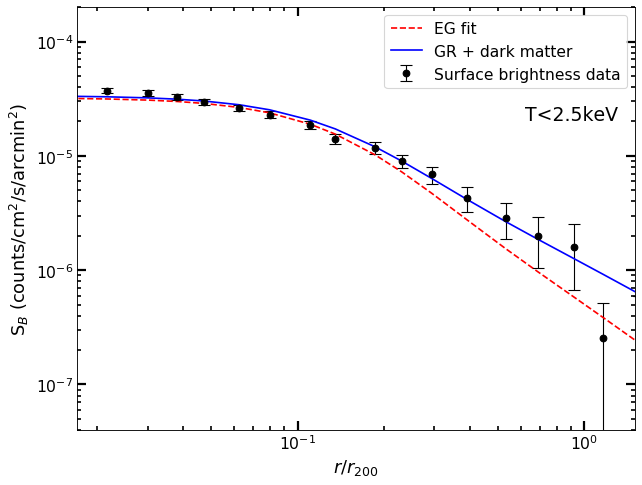}
\caption{X-ray surface brightness fits}
\label{fig:mainA}
\end{subfigure}%
\begin{subfigure}{.5\linewidth}
\centering
\includegraphics[width=.85\textwidth]{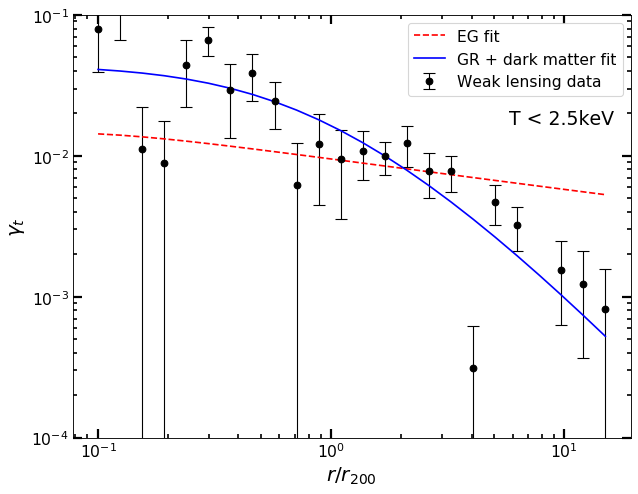}
\caption{Weak lensing fits}
\label{fig:mainB}
\end{subfigure}\\[1ex]
\begin{subfigure}{0.5\linewidth}
\centering
\includegraphics[width=.85\textwidth]{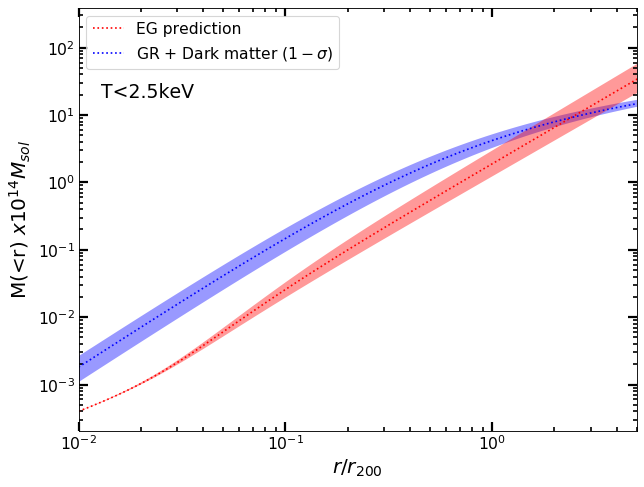}
\caption{Mass distributions in the two models}
\label{fig:mainC}
\end{subfigure}
\begin{subfigure}{.5\linewidth}
\centering
\includegraphics[width=.85\textwidth]{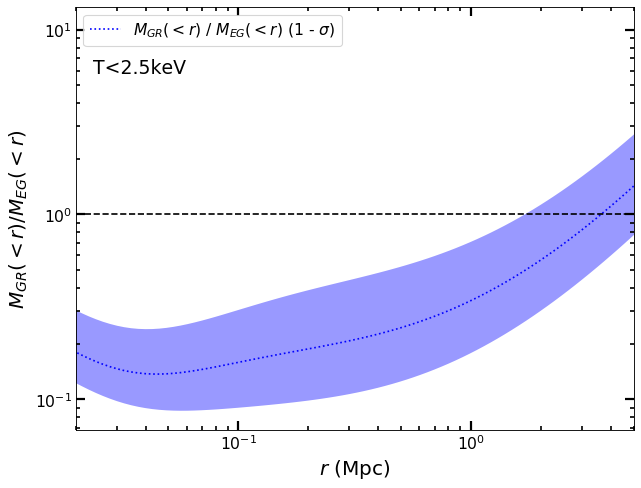}
\caption{Ratio of the mass distributions}
\label{fig:mainD}
\end{subfigure}
\caption{A comparison of the Emergent Gravity and the standard model (GR + dark matter described by the NFW profile) results (for the $T < 2.5$ keV bin roughly corresponding to galaxy groups). Figure $a$ shows the surface brightness fit for both models. Figure $b$ shows the weak lensing (tangential shear) fit for both models. The mass profiles were calculated using the best-fit parameters, with the blue and red bands corresponding to the 1-$\sigma$ confidence intervals. The goodness of fit statistics are summarized in table \ref{cluster_stack_goodness_of_fit}. }
\label{fig:main_results2}
\end{figure}

\section{Results and Discussion}
\label{results}

The Coma cluster results above indicate that Emergent Gravity is capable of producing fits that are generally comparable to the standard model fits and are in agreement with the observational data (within the shown uncertainties). The best-fit parameters from the gas temperature and the weak lensing data then result in mass distributions for the two models that are in agreement for $250$ kpc $< r <$ 700 kpc. The calculated mass distributions can be compared with other results in the literature, such as \cite{brownstein}, where the total mass profile was determined using X-ray data or \cite{lokas}, where elliptical galaxy velocity moments were used instead. In general, our GR + dark matter profile, within the given uncertainties, is in good agreement with the mentioned results from the literature, with the exception of around $r\sim 1$ Mpc, where the profiles in the mentioned papers fall between our Emergent Gravity and GR results. Overall this indicates that the Emergent Gravity result underestimates the total mass distribution for $r \lesssim 250$ kpc and overestimates it for $r \gtrsim 800$ kpc given 1-$\sigma$ confidence intervals. The values in table \ref{coma_goodness_of_fit} also indicate that, despite requiring more free parameters, GR is still the preferred model with lower $\chi^{2}$ and $BIC$ values. 

\begin{table}[ht!]
\centering
\begin{tabular}{lcccl}
                    & \textbf{$\chi^{2}$} & \textbf{Data points} & \textbf{D.o.f} & \textbf{BIC} \\ \hline \hline
\textbf{GR (bin 1):\,\,\,\,\,\,\,\,\,\,\,\,} &   98                                                  &    41                    &  6               &   59           \\
\textbf{GR (bin 2):\,\,\,\,\,\,\,\,\,\,\,\,} &      199                                                &   39                     &    6            & 79            \\ 
\textbf{EG (bin 1):\,\,\,\,\,\,\,\,\,\,\,\,} &      414                                                &   41                     &    4            & 109              \\ 
\textbf{EG (bin 2):\,\,\,\,\,\,\,\,\,\,\,\,} &      658                                                &   39                     &    4            &  122              \\ \hline
\end{tabular}
\caption{Goodness of fit statistics. $BIC$ is the Bayesian information criterion statistic. The statistics are combined for both the X-ray temperature and the weak lensing fits. See section \ref{results} for a discussion of these values.}
\label{cluster_stack_goodness_of_fit}
\end{table}

Figure \ref{fig:main_results} shows the results for the clusters with temperatures higher than 2.5 keV, which roughly corresponds to galaxy clusters (rather than galaxy groups). In this case the surface brightness fits are comparable for both models. However, the tangential shear profile fit in Emergent Gravity is significantly worse than the corresponding GR result. In general we found that Emergent Gravity could not simultaneously fit both datasets with accuracy. In other words, if we want to fit the surface brightness profiles accurately for $r/r_{200} \lesssim 2 \times 10^{-1}$, the resulting tangential shear profile will have a gradient that is too large to agree with the observational data for large values of $r$. This results in the total mass distributions in Emergent Gravity and GR that agree only at around $r \gtrsim  800$ kpc. 

In figure \ref{fig:main_results2}, for the $T < 2.5$ keV objects (roughly corresponding to galaxy groups) a similar trend emerges. In this case, the Emergent Gravity tangential shear fits are even poorer resulting in total mass distributions that agree well only for $r \gtrsim 1.5$ Mpc. Otherwise, for $r \gtrsim 10$ Mpc, the EG fits are in strong tension with the data. The values in table \ref{cluster_stack_goodness_of_fit} also indicate that GR is strongly preferred. Also, it is important to note that the rather high values of the $\chi^{2}$ are dominated by the contribution from the outlier points at $r/r_{200} \approx 2.7$, $r/r_{200} \approx 5.0$ for bin 1 and $r/r_{200} \approx 4.0$ for bin 2. However, removing the outlier points does not lead to a different conclusion regarding the preferred model. 

The results in figures \ref{fig:main_results} and \ref{fig:main_results2} are in agreement with most of the other results in the literature.  Specifically, in \cite{ettori} X-ray and SZ effect data is used to deduce the baryonic and, in turn, the total mass distributions for Emergent Gravity and the standard model, resulting in distributions very similar to ours (see figure 3 in \cite{ettori} in particular). More recently, in \cite{ettori2} the same approach was extended for a larger sample of clusters, once again resulting in mass distributions that agree only at around 1 Mpc radial scales. In \cite{hodson} the Emergent Gravity scaling relation is used to calculate acceleration radial distributions, again resulting in profiles for GR and Emergent Gravity, that only become comparable for $r \gtrsim 2$ Mpc. Finally, the results reported in \cite{halenka_miller} are closer to our results for the Coma cluster. Note, however, that such comparisons with other results in the literature should be treated with caution, as the methods and the datasets used to derive them are in general distinct and are affected by different systematics. 

Comparing the results for the galaxy clusters, groups and the Coma cluster indicates that, in general, Emergent Gravity seems to work better for massive clusters. This in turn means that accurate measurements of the total galaxy mass distribution (which dominates over the intracluster gas mass at low radii and hence could push the predicted mass profiles closer to those predicted in GR), are of special importance. In order to test the importance of the stellar galaxy mass measurements on our final results, we repeated the analysis outlined above, for the 58 cluster stack with various $M_{gal}(r)$ distributions (which were compared with the galaxy mass distribution of the Coma cluster). The results in appendix \ref{appendix:c} (fig. \ref{fig:a1}) indicate that having larger galaxy masses shifts the Emergent Gravity results closer to the GR results making them increasingly more similar to the Coma cluster result as expected. 
Finally, we investigated how the results were affected by relaxing some of the assumptions in the derivation of the scaling relation in eq. (\ref{EG}). In particular, as the authors point out in \cite{halenka_miller}, eq. (\ref{EG}) originally comes from an inequality between the left and right hand sides, which are ultimately set to be equal. Hence, they propose a phenomenological model, in which the $r^{2}$ term in the numerator of eq. (\ref{EG}) is replaced by $r_{a}r$, where $r_{a}$ is a constant. Analysis in \cite{halenka_miller} shows that $r_{a} = 1.2$ Mpc leads to a good agreement between the EG prediction and the data. In appendix \ref{appendix:c} fig. \ref{fig:a2} we repeated our analysis using the mentioned toy model and found that it does indeed lead to a significantly better agreement between the GR and EG models.

In summary, the model of Emergent Gravity offers a unique perspective in modifying General Relativity. Even though the model, in its original form, is in tension with the presented data, any conclusion on the ultimate validity of the model is rather premature; to fully evaluate the model, extensive theoretical and observational work is required. In particular, to account for the effects of high redshift clusters, the model must be further developed to take into account the variations of the Hubble parameter with redshift. Similarly, field and geodesic equations need to be derived to fully account for the lensing effects. 

In this regard, the recently developed Hossenfelder's interpretation of EG is very promising as it re-expresses the model in a more conventional form and allows one to deal with more general mass distributions \cite{hossenfelder}. This covariant formulation allows writing down a Lagrangian, which offers multiple advantages, such as easily comparing the theory with the many other scalar, vector and tensor theories of modified gravity as well as deriving the field equations and other predictions in the usual manner (see the appendix \ref{appendix:a} for a wider discussion of the properties of the Lagrangian). Finally, having a Lagrangian formulation would in principle allow deriving a geodesic equation and the corresponding weak lensing equations, which would be the next natural step in developing tests of Emergent Gravity. 
 
On the observational side, a natural extension of the work presented here would be expanding the cluster stack using the newest data releases from the Dark Energy Survey \cite{des}. With hundreds of galaxy clusters available to stack, we could very significantly improve the constraints on the model. In addition, having a large sample of clusters of different shapes and redshifts would allow a more extensive analysis of the effects of systematics on the final results. Having more accurate measurements of all the cluster mass components (e.g. stellar galaxy masses) would also significantly reduce the uncertainties. Finally, in future work we also plan to explore galaxy cluster simulations, which would allow free exploration of mass distributions of different sizes and shapes. This will allow us to test assumptions made in this work more rigorously.

\acknowledgments

This work would not be possible without the stacked galaxy cluster profiles constructed by H. Wilcox using the data from CFHTLenS and the XCS XMM-Newton cluster surveys. In addition, discussions with C. Miller and V. Halenka have been very beneficial in deducing the baryonic mass distributions for the Coma cluster and the cluster stack and, in general, getting the research project started. Help from K. Romer was very important in understanding the properties of our cluster stack data and comparing our results with the previous research in the literature. KK is supported by the European Research Council under the European Union's Horizon 2020 programme (grant agreement No.646702 ``CosTesGrav"). DB, KK and RCN are supported by the UK STFC grant ST/N000668/1. 
\\







\providecommand{\href}[2]{#2}\begingroup\raggedright\endgroup







\begin{appendices}

\section{Hossenfelder's Formulation of Emergent Gravity}
\label{appendix:a}

In the recent covariant formulation of Emergent Gravity, the displacement field $u(r)$ is treated as an extra vector field that originates from the volume term in the total entropy equations in Verlinde's theory. The vector field \textbf{u} couples to baryonic matter and drags on it to create an effect similar to dark matter. The proposed Langragian for the vector field \textbf{u} is given by: 

\begin{equation}
    \mathcal{L} = M_{pl}^{2}R + \mathcal{L}_{M} - \frac{u^{\mu}u^{\nu}}{Lu}T_{\mu \nu} + \frac{M_{pl}^{2}}{L^{2}} \chi^{3/2} - \frac{\lambda^{2}M_{pl}^{2}}{L^{4}} (u_{\kappa}u^{\kappa})^{2},
    \label{lagrangian}
\end{equation}

\noindent with $\mathcal{L}_{M}$ as the matter Lagrangian, $L$ as the Hubble radius, $T_{\mu \nu}$ as the stress-energy tensor, $\chi$ as the kinetic term of the vector field and $\lambda$ as the mass term\footnote{Note that here we use the correct form of the Lagrangian from \cite{stojkovic2}, which also contains an illuminating discussion of the stability of the de Sitter space solution of eq. \ref{lagrangian}.}.

The form of the Lagrangian indicates some subtle differences between the Verlinde and Hossenfelder formulations. In particular, the extra terms in the Lagrangian indicate that even when no mass is present in the system, the field \textbf{u} does not vanish. Or, in other words, stress-energy conservation would require the field \textbf{u} to be a source of gravity as well. This means that the solutions for the total potential, in general, will not be identical to those derived by Verlinde and will contain correction terms. Another interesting feature of the Lagrangian is the 2/3 power of the kinetic term. There have been a number of modified gravity approaches that have a similar kinetic term, most notably \cite{berezhiani}, where a theory of dark matter superfluidity is proposed. As discussed in section 5.1 in \cite{hossenfelder}, the Langrangian above can be solved for $\phi = \sqrt{-u^{\alpha}u_{\alpha}}/L$, however the solution contains an integration constant that cannot be determined analytically and would require numerical solutions. Finding these solutions is out of the scope of this work; however, further exploration of the covariant formulation of EG for spherical and non spherical mass distributions, and comparison of the results with the predictions in Verlinde's original formulation, will be an interesting direction for future work.

\section{NFW Weak Lensing Equations}
\label{appendix:b}

Here we outline the full tangential shear equations under the assumption that the dark matter in a given cluster is distributed according to the NFW profile \cite{nfw1}: 

\begin{equation}
    \gamma_{NFW}(x) = \begin{cases} 
                                    \frac{r_{s}\delta_{c}\rho_{c}}{\Sigma_{c}} g_{<}(x) & (x < 1) \\ 
                                    \frac{r_{s}\delta_{c}\rho_{c}}{\Sigma_{c}} \Big( \frac{10}{3} + 4\ln{(\frac{1}{2})} \Big) & (x = 1)\\ 
                                    \frac{r_{s}\delta_{c}\rho_{c}}{\Sigma_{c}} g_{>}(x)  & (x > 1),
                      \end{cases}
                      \label{nfw_shear1}
\end{equation}

\noindent where we defined $x = r/r_{s}$ with $r_{s}$ as scale radius, $\rho_{c}$ as the critical radius and $\delta_{c}$ is the characteristic overdensity of the dark matter halo. $g_{>}(x)$ and $g_{<}(x)$ are given by: 

\begin{equation}
    g_{<}(x) = \frac{8 \arctan(\sqrt{(1-x)/(1+x)})}{x^{2}\sqrt{1-x^2}} + \frac{4}{x^2} \ln{(x/2)} - \frac{2}{(x^{2}-1)} + \frac{4 \arctan(\sqrt{(1-x)/(1+x)})}{(x^{2} -1)(1-x^{2})^{1/2}     } 
\label{nfw_shear2}
\end{equation}

\begin{equation}
    g_{>}(x) = \frac{8 \arctan(\sqrt{(1-x)/(1+x)})}{x^{2}\sqrt{x^2-1}} + \frac{4}{x^2} \ln{(x/2)} - \frac{2}{(x^{2}-1)} + \frac{4 \arctan(\sqrt{(1-x)/(1+x)})}{(x^{2}-1)^{3/2}     } 
\label{nfw_shear3}
\end{equation}

\section{Results with Different Galaxy Mass Distributions and a Modified Scaling Relation}
\label{appendix:c}

\begin{figure}[!ht]
\centering
\captionsetup[subfigure]{justification=centering}
  \begin{subfigure}[b]{0.49\textwidth}
    \includegraphics[width=\textwidth]{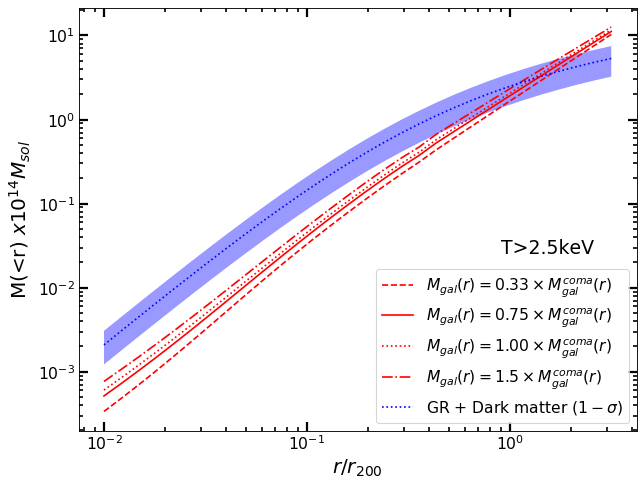}
    \caption{\small Results with different $M_{gal}(r)$}
    \label{fig:a1}
    
  \end{subfigure}
  \begin{subfigure}[b]{0.49\textwidth}
    \includegraphics[width=\textwidth]{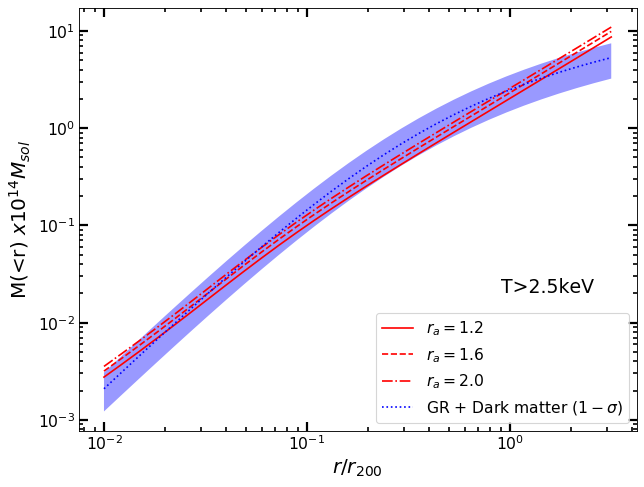}
    \caption{Results with modified $M_{D}(r)$}
    \label{fig:a2}
  \end{subfigure}
  \caption{Analysis of how the main results are affected by varying the galaxy mass function (compared to the Coma cluster) and modifying the scaling relation for $M_{D}(r)$.}
  \label{fig:a}
\end{figure}

As shown in eq. (\ref{baryonic_mass}) the total baryonic mass distribution was determined by combining the contributions from the stellar galaxy masses and the intracluster gas. Since the stellar mass component could not be determined from our cluster stack data, we instead used the SDSS (for the Coma cluster) and the CFHTLenS (for the cluster stack) open access data catalogues to determine it for each cluster individually. Since the stellar galaxy mass component dominates the total cluster mass at small radii, it is important to investigate the effects of underestimating/overestimating it. In addition, one of the possible reasons of why EG works relatively better with the Coma cluster data (see fig. \ref{fig:fig:mean and std of net44} and \ref{fig:mainC}) could be that the cluster is known to be unusually massive. Hence, here we present the total mass distributions deduced in the same way as in fig. \ref{fig:main_results}, but now with a galaxy mass distribution closer to that of the Coma cluster (i.e. being equal to $0.33-1.5 \times M_{gal}^{Coma}(r)$, where $M_{gal}^{Coma}(r)$ is the Coma galaxy mass distribution determined from the SDSS data). As fig. \ref{fig:a1} illustrates, having significantly larger galaxy masses (while keeping the intracluster gas component unchanged) results in a better agreement between the standard model (GR + cold dark matter), Emergent Gravity and the observational data. 

It is also important to point out that the scaling relation in eq. (\ref{EG}) originally comes from the following inequality:

\begin{equation}
    \int^{r}_{0}\epsilon_{D}^{2}(r')A(r')dr' \leq V_{M_{B}},
    \label{strain2}
\end{equation}

\noindent which puts a bound on the maximum value of $\epsilon_{D}(r)$ caused by some baryonic matter distribution $M_{B}(r)$ (see section 7.1 in \cite{b} for more detailed discussion). In the original derivation, the maximum value for the left hand side of eq. (\ref{strain2}) is chosen, however, as discussed in \cite{tortora} and \cite{halenka_miller}, different values can arguably be considered. If we introduce an extra parameter $r_{a}^{2}$ on the right hand side of eq. (\ref{strain2}), we can consider baryonic matter causing a different amount of maximum strain $\epsilon_{D}$. The modified inequality then results in the following altered scaling relation: 

\begin{equation}
    M_{D}^{2}(r) = \frac{cH_{0}r_{a}r}{6G} \frac{d(M_{B}(r)r)}{dr},
\label{mod_EG}
\end{equation}

\noindent where $r_{a}$ is a parameter that describes how the elastic medium in Verlinde's theory is affected by the baryonic matter. If we use the modified scaling relation and carry out our analysis again, the results in fig. \ref{fig:a2} are obtained. In agreement with the results in \cite{halenka_miller} and \cite{tortora}, for values of $r_{a} \approx 1.2$ we find a good agreement between GR and EG. 

\end{appendices}

\end{document}